# TDPAC and first-principles study of electronic and structural properties of Pd-vacancy complex in undoped germanium

Adurafimihan A. Abiona[*], Williams Kemp and Heiko Timmers

*School of Physical, Environmental and Mathematical Sciences, The University of New South Wales, UNSW Canberra, PO Box 7916, Canberra BC 2610, Australia*

**Abstract**

Pd is one of the metals suitable for inducing low-temperature crystallization in Ge. However, it is not clear how residual Pd atoms are integrated into the Ge lattice. Therefore, time-differential γ-γ perturbed angular correlations (TDPAC) technique using the $^{100}$Pd($\to^{100}$Rh) nuclear probe produced by recoil implantation has been applied to study the hyperfine interactions of this probe in single-crystalline undoped Ge. A Pd-vacancy complex aligned along the <111> crystallographic direction with a unique interaction frequency of 8.4(5) Mrad/s has been identified. This complex was measured to have a maximum relative fraction of about 76(4)% following annealing at 350 $^{o}$C. Further annealing at higher temperatures reduced this fraction, possibly via dissociation of the complex. Calculations suggest dissociation energy of 1.94(5) eV for the complex. DFT calculations performed in this work are in reasonable good agreement with the experimental values for the electric-field gradient of the defect complex in Ge. The calculations predict a split-vacancy configuration with the Pd on a bond-centred interstitial site having a nearest-neighbour semi-vacancy on both sides (V-Pd$_{BI}$-V).

# 1. Introduction

Metal-induced crystallization (MIC) has been explored for the processing of complementary metal-oxide-semiconductor (CMOS) technology based on germanium (Ge). CMOS integrated circuits are advantageous because they feature low noise and static power. Palladium (Pd) has been shown to induce MIC at a lower temperature than most metals; therefore it is of a particular interest [1, 2]. The device performance, however, will depend on the concentration and the lattice location of residual Pd atoms following the crystallization. Furthermore, the structural integration of Pd atoms into the host lattice and the possible formation of defect complexes may have an impact on the electronic properties of the material. It is well known that transition-metal (TM) impurities in elemental semiconductors – silicon (Si) and Ge – exist in multiple deep-level states with different charges, and they are efficient charge-carrier traps [3]. Hence, the presence of these metal impurities in Ge might degrade or enhance the desired performance of Ge-based devices even in small quantities [4]. For instance, Pd, Pt and Au have been used to control the lifetime of fast-switching diodes [5]. Recent research has also shown that TM's, in particular hafnium, as a co-dopant can retard the diffusion of major dopants like phosphorus in Ge [6].

In the last three decades, hyperfine experimental techniques, such as nuclear Mössbauer spectroscopy, quadrupole resonance, nuclear magnetic resonance and time-differential perturbed angular correlations (TDPAC) [7], have been widely used to study materials from the viewpoint of solid-state physics, chemistry and biology in order to explain the microscopic environments of constituents or impurities and the nature of chemical bonding in different kind of molecules and compounds. Of special interest to this study is TDPAC because it is possible to produce the probe $^{100}$Pd($\rightarrow^{100}$Rh), which is well suited to study the local defect dynamics of Pd atoms in Ge. TDPAC is a very sensitive tool for investigating the electric-field gradient (EFG) created by non-cubic symmetry at the probe site as result of a defect in its vicinity. Specifically, the magnitude of the strongest component

$V_{zz}$ of the EFG tensor can be derived from the quadrupole interaction frequency $v_Q$, which is one of the measurable parameters in TDPAC spectroscopy. Such information could give the configurations, the formation and the dissociation of Pd defect complexes as well as determine their orientation within the host lattice.

In this paper, the EFG, the second derivative of the electrostatic potential with respect to the spatial coordinates, has been used to investigate Pd-defect complexes in undoped Ge. EFG is a quantity which is extremely sensitive to the symmetry of the electronic-charge density. It may be determined through its hyperfine interaction effects with the nuclear quadrupole moment $Q$ of a suitable probe nucleus introduced during TDPAC measurement. Due to the $r^{-3}$ dependence of the EFG, it is very sensitive to even the most subtle changes at a subatomic scale of the electronic charge density close to the probe nucleus. Consequently, it is a powerful parameter for extracting structural information, such as lattice defects and lattice disorders, or electronic information, such as chemical bonds, in solids [8-10]. Thus, the EFG reflects the bonding of the probe in its atomic neighbourhood. Essentially, the nearest neighbours of the probe contribute to the EFG.

In the last few decades, electronic-structure calculations, in particular density-functional theory (DFT), have become essential in supporting and improving experimental results [8-11]. The interpretation of TDPAC data and assigning a particular configuration and direction (sign) to the EFG of the defect complex is very complex. To complement the hyperfine properties measured by TDPAC, DFT calculations are often performed. These calculations can predict the sign of the EFG tensor $V_{zz}$ and often give information about the configuration of the complex and its electronic properties. Therefore, the experimental work has been extended here and DFT calculations were performed to better understand the electronic and structural properties of the Pd-defect complex in Ge. The calculations were carried out using the gauge-including projector-augmented wave (GIPAW) method [11-13], as implemented in the plane-wave code Quantum Espresso (QE) [14]. Similar calculations were also performed for Si. These allow a comparison with Ge, and enable a broader interpretation of the results. Some calculations were also performed for Rh, the daughter in the $^{100}$Pd($\rightarrow^{100}$Rh) decay scheme, in Ge and Si to shed more light on the influence of Pd transmutation to Rh in Pd-defect TDPAC measurements.

This paper is organized as follows: section 2 gives concise details on nuclear production and implantation in undoped Ge, TDPAC spectroscopy and the subsequent data processing and analysis. In section 3, brief details on the computation are given. In sections 4 and 5, experimental and computational results are reported and discussed, respectively. Finally, in section 6, conclusions are presented.

## 2. Experimental details

High-purity Czochralski-grown single-crystal wafers with a (111) polished plane of undoped Ge were studied. The 14UD Pelletron accelerator at the Australian National University was used to synthesize the $^{100}$Pd($\rightarrow^{100}$Rh) probe using the $^{92}$Zr ($^{12}$C, 4n)$^{100}$Pd fusion-evaporation reaction. A 70 MeV $^{12}$C$^{5+}$ beam was incident on a 2.5-μm-thick zirconium target over 20 hours at a beam current of about 1 μA [15]. Along with other reaction products, $^{100}$Pd($\rightarrow^{100}$Rh) probes were directly recoil-implanted into the Ge samples. The probe ions had energies of up to 8 MeV. The samples were placed several millimetres outside the beam path beyond a scattering angle of 2°, as shown in figure 1. This drastically reduces the number of scattered-beam atoms that are co-implanted into the samples, albeit significantly deeper than the probe atoms. The as-implanted Ge samples were annealed for 30 min at temperatures of 250 °C, 300 °C, 350 °C, 400 °C, 550 °C and 700 °C. The annealing was carried out in flowing nitrogen gas inside a tube furnace. Each annealing step was followed by TDPAC measurements. All measurements were carried out at room temperature.

The $^{100}$Pd nucleus decays with a half-life of $t_{1/2}$ = 3.6 d to the excited states of $^{100}$Rh (figure 2). The sequence of the excited states of $^{100}$Rh ($1^+ \rightarrow 2^+ \rightarrow 1^-$) are connected via an 84–75 keV γ-γ cascade [16]. The intermediate $2^+$-state has a nuclear quadrupole moment of $Q = 0.151(8)$ b [17] and is thus well suited for TDPAC measurements. The interaction between the nuclear moment of the intermediate state of $^{100}$Rh and the surrounding electromagnetic field provides information about the environments of the probes. EFG characterizes the probe environment. The EFG can be completely explained by measurement of the quadrupole coupling constant $\nu_Q$ and the asymmetry parameter $\eta$

$$\nu_Q = \frac{eQV_{zz}}{h} \qquad \eta = \frac{V_{xx}-V_{yy}}{V_{zz}} \text{ (with } 0 \leq \eta \leq 1\text{)}. \qquad (1)$$

$V_{zz}$ is the largest component of the diagonalized field tensor of the EFG in the principal axis system.

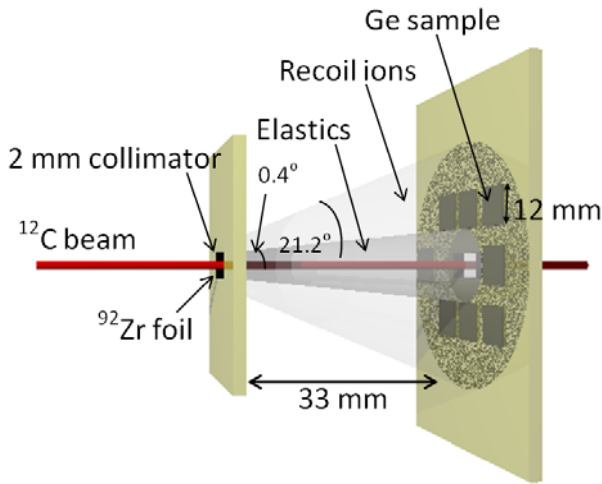

**Figure 1.** A schematic expanded view of the reaction chamber for the production and recoil implantation of the $^{100}$Pd($\rightarrow^{100}$Rh) TDPAC nuclear probes showing the samples, the beam and the spread of the elastics and the recoiled ions.

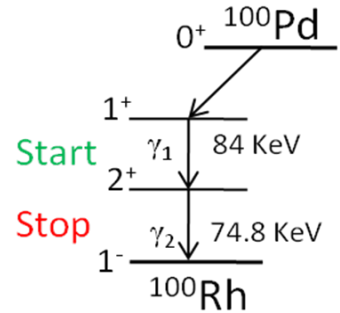

**Figure 2.** The partial decay scheme of $^{100}$Pd.

In this work, a standard slow-fast pulse TDPAC spectrometer with four BaF$_2$ detectors in the conventional planar geometry oriented at angle 90° to each other was used to characterize the EFG. Detected 84 and 75 keV γ-rays that overlapped in the energy spectrum provided start or stop signals to a time-to-analogue converter. Coincidence-time distributions $N(\theta, t)$ of the two γ-rays were measured for all detector combinations using NIM-standard electronics. The measured time spectra were corrected for statistical background events. The distributions for the 90° and 180° detector-angle combinations (with four combinations each) were geometrically averaged by their combinations as $\overline{N(\theta, t)}$. From the averaged spectra, the time differential anisotropy expressed as a ratio function *R(t)* was computed according to

$$R(t) = 2 \cdot \frac{\overline{N(180^0,\ t)} - \overline{N(90^0,\ t)}}{\overline{N(180^0,\ t)} + 2\,\overline{N(90^0,\ t)}} \tag{2}$$

and were fitted with least-squares function with an appropriate perturbation function as:

$$R(t) = A_{22} \sum_i f_i G^i_{22}(t), \qquad \sum_i f_i = 1 \tag{3}$$

Generally, different defect environments exist in the lattice, and this results in different perturbation functions $G^i_{22}(t)$ for each interaction, having fractional population $f_i$ of the probe nuclei, which can be treated with a linear superposition as:

$$G^i_{22}(t) = \sum_{n=0}^{10} S^i_n \cos[g_n(\eta)\omega^i_0 t] \exp[-g_n(\eta)\omega^i_0 \delta^i t]. \tag{4}$$

$\omega_0$ is the fundamental precession frequency of the intermediate angular momentum, $\omega_0 = 3v_Q/12$. The value of $\eta$ can be deduced from the frequency factors $g_n(\eta)$. In the exponential term, $\delta^i$ measures the average width of the distribution of the frequency around the mean value. The coefficients $S^i_n$ contain information about the geometry of the EFG with respect to the host lattice. Details of the TDPAC method are available in many textbooks [16, 18].

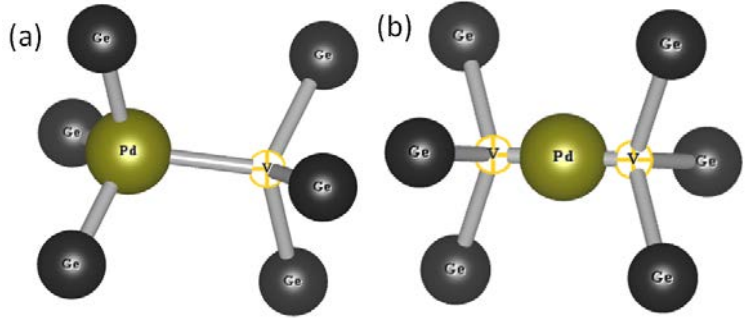

**Figure 3.** Two possible microscopic configurations of a Pd-defect complex in Ge: (a) the Pd-vacancy configuration with the impurity (Pd) on a substitutional (SS) site and a vacancy (V) as nearest neighbour ($Pd_{SS}$-V); (b) the split-vacancy configuration with the impurity (Pd) on a bond-centred interstitial (BI) site with a nearest-neighbour semi-vacancy (V-$Pd_{BI}$-V) on both sides.

## 3. Computational details

Theoretical calculations were performed with the QE package using the GIPAW method for the EFG calculations [14]. QE is a plane-wave code that uses first principles application of density functional theory (DFT), and the GIPAW method ensures the reconstruction of all-electron wavefunctions with computationally efficient pseudopotentials. This approach also permits the study of the effects of atomic relaxations induced by the probe atom on the hyperfine parameters, as well as the electronic and structural properties of the host lattice.

Ultrasoft Vanderbilt pseudopotentials were used for Ge, Pd and Rh, while a norm-conserving pseudopotential was used for Si, including the nonlinear core correction in the exchange-correlation terms [19]. Scalar relativistic pseudopotentials were used; but, the spin-orbit interaction was not included. The exchange-correlation energy effects were treated using the generalized gradient approximation (GGA) of Perdew-Buurke-Ernzerhof [20]. The calculations were performed with a cut-off energy of 45 Ry and 50 Ry for Ge and Si supercells, respectively, with a uniform grid of $k$ points in 4×4×4 Monkhorst and Pack grid [21] which are required for convergence, in the diamond cubic cell. The optimized lattice constants calculated for the bulk Ge and Si were 5.762 Å and 5.469 Å, respectively, which are in good agreement with the experimental values of 5.646 Å and 5.405 Å, within the tolerance 3 % of DFT calculations [22].

The Pd-defect complex was introduced into the supercell by substituting a Pd atom for a Ge or Si atom and removing a nearest-neighbour Ge or Si atom to create a vacancy (V). The complex was placed in the middle of the 64-atom supercell, which was used for the calculations. The symmetries of Ge and Si were lowered from $T_d$ to $C_{3v}$ when the Pd-defect complexes were introduced. The introduction of a TM into Ge and Si introduces some deep-level states into these semiconductors and causes the complexes formed by the metal with other defects to have a net charge state. To account for this charge state in the calculations, additional electrons or holes were introduced to the system, and compensated with a uniform charged background [23]. As a result, the system becomes more metallic; therefore, the wavefunctions were smeared using Gaussian functions with a very narrow smearing width of 0.0025 Ry and 0.004 Ry for Ge and Si, respectively.

In this work, two possible microscopic configurations for a Pd-defect complex in Ge and Si were investigated, as shown in figure 3. In the first configuration [figure 3(a)], the Pd atom occupies a substitutional (SS) site and pairs with a nearest-neighbour vacancy ($Pd_{SS}$-V) attributed to a Pd-V complex in Si [24, 25]. For this configuration, the calculations were performed without any relaxation of the atoms to preserve the configuration. The second configuration [figure 3(b)] is the split-vacancy configuration with the Pd on a bond-centred interstitial (BI) site pairing with a nearest-neighbour semi-vacancy (V-$Pd_{BI}$-V) on both sides [10, 26]. For this configuration, the calculations were started with a Pd atom on an SS site, and allowed the six nearest-neighbour Ge or Si atoms and the Pd atom to fully relax, using the conjugate-gradient method to minimize the total energy until all residual forces on atoms are less than 1 mRy/a.u.

In GIPAW formalism, the calculation of the EFG is implemented in two stages: the pseudopotential generation and the EFG tensor calculation. The all-electron response is reconstructed from the pseudopotential during its generation by using Blöchl's PAW method [27] as was explained by Petrilli *et al.* [12]. For more details on this implementation, see the work of Profeta *et al.* [11], which gives a comprehensive description, and the review of Charpentier [28].

## 4. Experimental results and discussion

The modulations of the *R(t)'s* were unique, so that two probe fractions were assumed for all the fits. A typical *R(t)* for a Pd-defect complex in Ge annealed at 350 °C. The first probe fraction represents Pd-defect pairs that are in a unique and otherwise undisturbed Ge lattice environment. The corresponding contribution to *R(t)* is thus a pronounced modulation without damping. The second probe fraction includes all other probe environments. Since a variety of probe environments may occur, the contribution of this fraction is a heavily damped function. The fits performed typically achieved $\chi^2 = 1.7$ or better.

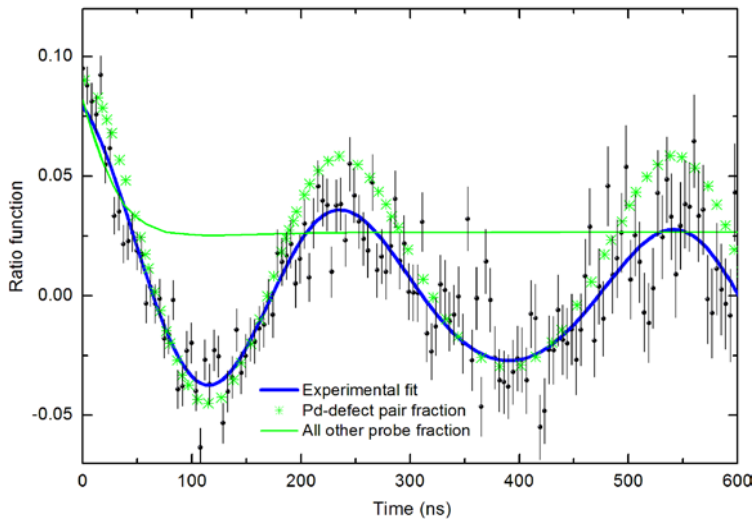

**Figure 4.** (Color online) A typical ratio function R(t) data (solid sphere with error bars) fitted with two fractions. The modulated green curve (*) represents the site-located probes pairing with nearest-neighbour vacancies – a unique hyperfine interaction. The damped flat green curve (thin line) represents the remaining probes in a variety of lattice environments – an exponentially damped contribution. The sum of the two fractions is represented by the blue modulated curve (bold line).

The data were fitted with a unique interaction characterised by an interaction frequency $\omega_0$ = 8.4(5) Mrad/s and an asymmetry parameter $\eta$ = 0. This corresponds to a quadrupole coupling constant of $v_Q$ = 10.7(5) MHz. This quadrupole coupling constant is close to the 13.1(1) MHz observed for the same $^{100}$Pd($\rightarrow^{100}$Rh) probe in highly doped n-type

Si [24], where it was attributed to the formation of Pd-vacancy pairs, with the Pd being substitutional and the vacancy (V) located in the <111> crystallographic direction. This directionality was inferred from TDPAC orientation measurements [25]. This suggests that the two effects are of similar origin. In addition to this observation in Si, the only possible defect in ion-implanted undoped Ge is a vacancy. We attribute the observed defect complex to a Pd-V complex pairing along the <111> crystal direction.

Such probe pairing with a vacancy in doped Ge has also been reported for TDPAC measurements with an $^{111}$In($\rightarrow^{111}$Cd ) probe [29, 30]. These authors interpreted this observation as consisting of two types of defect complexes, substitutional Cd pairing with the nearest-neighbour vacancy and interstitial Cd pairing with the nearest vacancy, respectively. Both complexes were also found to be oriented along the <111> crystal direction.

As described earlier, different research groups have reported different results, and the configuration of this TM-defect complex is still an open question. In order to gain more insight into this configuration, DFT calculations of the EFG's are often done. Full details and results are given in section 5.

A sequence of TDPAC measurements were performed on the sample after annealing at different temperatures, as shown in figure 5. The measured data were fitted with a least-squares function as described earlier, with the unique interaction frequency $\omega_0$ = 8.4(5) Mrad/s. The modulation patterns are very similar, having the first minimum approximately at $t$ = 100 ns and the second at $t$ = 380 ns for all the annealing temperatures. This indicates that the same defect complex is present at all the annealing temperatures. The effects of implantation damage are apparent in 5(a) as a damping of *R(t)*. This is caused by random breaking of the lattice symmetry at different probe sites, resulting in a finite distribution of EFG's acting on the ensemble of probes. Figures 5(b), (c) and (d) show a gradual recovery of the lattice, with a unique probe environment becoming apparent as a pronounced modulation pattern at an annealing temperature of 350 °C.

The relative fraction of the Pd-V complex as a function of annealing temperature was studied. The maximum relative fraction of the complex experiencing the interaction frequency $\omega_0$ = 8.4(6) Mrad/s was about 76(4)% after annealing at 350 °C. Beyond this temperature, the relative fraction is reduced. In contrast, the fraction of all other probes in the other environments increased; this could be as a result of dissociation of the complex. From the fact that the Pd-V complex started dissociating at 350 °C, we can estimate the activation energy for the dissociation of the complex at this annealing temperature. A detailed explanation of the estimation can be found in the work of Skudlik *et al* [31] and Wahl *et al.* [32]. The activation energy of the dissociation is calculated from

$$E_a = K_B T_n \ln\left\{\frac{v_0 \Delta t}{N}\left(\ln\left[\frac{f(T_{n-1})}{f(T_n)}\right]\right)^{-1}\right\}, \qquad (8)$$

where $K_B$ is the Boltzmann constant, $T_n$ is the annealing temperature during dissociation and $v_0$ the attempt frequency, whose value is equal to the phonon frequency $v_0 \approx 10^{12}$ s$^{-1}$ [31, 32].

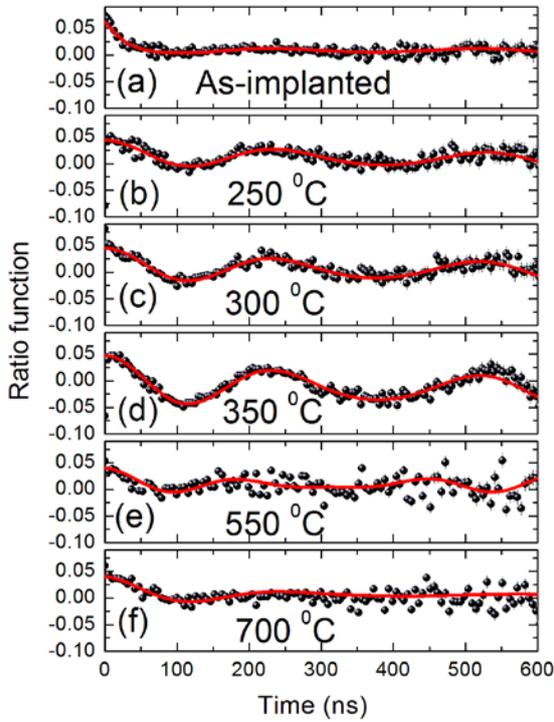

**Figure 5.** Measured TDPAC ratio functions, R(t), for $^{100}$Pd($\rightarrow ^{100}$Rh) in undoped Ge. The samples were annealed at the indicated temperatures before TDPAC measurements. Fits with a quadrupole coupling constant of 10.4(8) MHz assuming Pd pairs with the nearest-neighbour vacancy (V) along the <111> direction are shown.

$N$ is the number of times the vacancy is retrapped before escaping. $f(T_{n-1})$ and $f(T_n)$ are the initial and final fractions of the complex after the (n-1)th and the *n*th annealing, respectively. Using a one-step model without retrapping (i.e. $N = 1$), the estimated activation energy of dissociation at the annealing temperature of 350 °C with 300 °C for the annealing step before was 1.94(5) eV. This energy is near to the 2.5(7) eV that was estimated for similar defects in a Pd-V complex in Si [24] and the 1.64 eV in an In-V complex in Ge [33].

## 5. Computational results and discussion

5.1. Electric-field gradients and electronic-structure calculations

The simplest (and the traditional) approach to calculate EFG's is to use the point-charge model (PCM), in which the charges are assumed to be point-like and localized at the nucleus position. However, the PCM has many shortcomings, some of which are outlined in the work of Jalali Asadabadi *et al.* [34]. Most importantly, it is only suitable for complete ionic systems, for instance in metal oxides [35], in which the atoms have well-defined charge states. Therefore, it is not appropriate for elemental semiconductors in which bonding is predominantly covalent, with some degree of pure or half-metallic character and small ionic properties when doped [36]. Thus, for the system under investigation, the DFT method has been used for the calculations of EFG's.

The $^{100}$Pd($\rightarrow^{100}$Rh) radioactive probe used in this work decays from a $^{100}$Pd probe (half-life 3.6 days) to $^{100}$Rh (half-life 214 ns and spin 2). The parent isotope ($^{100}$Pd) determines the lattice location of the probe because it is the isotope that is actually implanted and traps other defects in its vicinity. The Pd-defect complex formed is converted to an Rh-defect after the decay. Although the defects are trapped by $^{100}$Pd isotopes, the $^{100}$Rh isotope gives information about the EFG in the probe's vicinity because the EFG is measured by detection of the γ-γ cascade emitted by the $^{100}$Rh isotope. After the decay, the $^{100}$Rh isotope remains in almost the same location as the parent isotope because the time-scale of the radioactive decay is an order of magnitude smaller than atomic diffusion in lattices [37]. Thus, the daughter isotope $^{100}$Rh is used for the EFG calculations.

To answer the question earlier raised about the configuration of the Pd-defect complex in the elemental semiconductors the EFG's were calculated. The calculated values of the largest component of the EFG tensors, $V_{zz}$, for different charge states of Rh-V complexes in Ge and Si are shown in Table II (in units of $10^{21}$ V/m$^2$) for the two possible configurations. The $V_{zz}$ at the Rh site for the experimental results were calculated from the quadrupole coupling constant $\upsilon_Q$ according to Eq. (1), with the nuclear quadrupole moment of the $^{100}$Rh probe taken as $Q = 0.151(8)$ b [17]. The EFG calculations for the two configurations are axially symmetric, i.e. $\eta = 0$, due to the constraints of the adopted $C_{3v}$ symmetry; this is in agreement with experimental observations in previous work in doped Si [24, 25], doped Ge [38] and this work in undoped Ge. TDPAC measurement does not give information on the direction (i.e. sign) of the $V_{zz}$ [9], therefore, the measured values are prefixed with "±" symbol to reflect this. Advantageously, DFT calculations complement the measurements in this regard by give the sign of the $V_{zz}$. Subsequently in this work, comparisons of the measured values with the calculated ones are done mainly on their magnitudes.

EFG calculations of Rh in Si were first performed to have a good understand of the systems since they are similar and less complicated to compute than that of Rh in Ge. The calculated value of $V_{zz} = -3.75 \times 10^{21}$ V/m$^2$ for a -1 charged state for the V-Rh$_{BI}$-V configuration is in close agreement with the extrapolated experimental value of $V_{zz} = 3.69(19) \times 10^{21}$ V/m$^2$ at 0 K for Si [24, 25]. Whereas, the calculated values of $V_{zz}$ for different charged states for the Rh$_{SS}$-V configuration from Brett *et al.* [24], Dogra *et al.* [30] and this work for Si, are not close to the experimental value (table 1). Similarly, the calculated values of $V_{zz} = -3.52 \times 10^{21}$ V/m$^2$ for a -2 charged state for the V-Rh$_{BI}$-V configuration are in reasonable close agreement with the experimental value of $V_{zz} = \pm 2.93(12) \times 10^{21}$ V/m$^2$ at room temperature, whereas the values for the Rh$_{SS}$-V configuration are not in agreement with the experimental value (table 1). These calculations confirm that the Pd-V complex in both Ge and Si have the V-Pd$_{BI}$-V configuration. The confirmation of the V-Rh$_{BI}$-V configuration in which an impurity atom occupies a BI site and pairs with two semi-vacancies, has already been observed by other researchers. Höhler *et al.* [10, 26] observed in their DFT calculations that most oversized impurities such as Cd occupy BI sites in the lattice of Ge and Si after full relaxation, in agreement with TDPAC measurements by other authors [39]. Watkins' studies on the Sn-vacancy pair in Si also confirmed this observation [40]. Likewise, Decoster *et al.* [41] through emission channelling and DFT calculations demonstrated that TM's do not only occupy SS sites in Ge, but also BI sites.

Table 1: Calculated EFG tensors $V_{zz}$ ($10^{21}$ V/m$^2$) at Rh sites, the sites for the substitutional (SS) Pd and bond-centred interstitial (BI) Pd in silicon and germanium for different charge states (q = 0, -1, -2). The experimental values at room temperature and the extrapolated value (in square parenthesis) for 0 K is also given for Si. Q = 0.151(8) barn is used for $^{100}$Rh [17].

| Charge | Si | | | Ge | |
|---|---|---|---|---|---|
| | $Rh_{SS}$-V[a] | $Rh_{SS}$-V | V-$Rh_{BI}$-V | $Rh_{SS}$-V | V-$Rh_{BI}$-V |
| 0 | +4.76 | +8.77 | -2.26 | +5.67 | -1.92 |
| -1 | +6.04 | +10.26 | -3.75 | +6.44 | -3.78 |
| -2 | - | +11.98 | -5.24 | +7.58 | -3.52 |
| Exp. | 3.58(19)[b] [3.69(19)] | | | 2.92(12)[c] | |

[a]Calculated values of references [29, 30], [b]Measured value of references [29, 30], [c]Measured value in this work.

After the configuration analysis, lattice relaxations of both the parent and daughter isotope impurities (i.e. Pd and Rh) were analysed to investigate the effects of the relaxation on the lattice-site locations of the impurities, which could have affected the defect complex configuration. Table 1 is a listing of the lattice relaxation of the Pd and Rh impurity atoms and one of the six nearest-host atoms (Ge or Si), computed as a percentage of the unrelaxed nearest-neighbour (NN) distance of the host for the V-$Pd_{BI}$-V and the V-$Rh_{BI}$-V configurations. In both Ge and Si, Pd and Rh moved towards the bond-centred site, which is halfway (50%) between the impurity and the vacancy in an unrelaxed supercell. Consequently, the six NN atoms contracted towards the impurity-vacancy complex; the relaxations are given in table 2 in the units of NN%. More specifically, the Pd atom moved 30% of the NN distance for the neutrally charged complex, while the Rh atom moved 25% of the NN distance for the neutrally charged complex in both Ge and Si. For the charged complexes, both Pd and Rh moved on average 47% of the NN distance in the two host materials. The similarity in the relaxations of the $^{100}$Pd parent and $^{100}$Rh daughter isotope atoms confirms that $^{100}$Rh remains in almost the same lattice location as the parent isotope after the transmutation. Therefore, the transmutation of the $^{100}$Pd isotope has a negligible influence on the lattice-site locations of the daughter isotope and effectually does not influence the defect configuration.

For more description of the defect-complex configuration, a detailed analysis of the electronic structure is done. The total density of states and partial density of states (PDOS) for Pd and Rh impurities in Ge are shown in figures 6(a) and (b), respectively. First, one can see that there is a similarity between the electronic states, in terms of the peak structures and the relative heights of the peaks in the PDOS in the two figures, which indicates that the introduction of the metal (Pd or Rh) impurities in Ge causes the same defect conditions. In essence, the defect complex and its chemical bonds formed in these materials as a result of these TM impurities are most likely the same, in agreement with the experimental and computational results in this work. Secondly, one can see that the largest electronic density at the Fermi level is due to the Ge $4sp^3$ hybridized state, with a large energy spread from the valence band maximum (VBM) to about 4 eV below the VBM. There are also some contributions from the $4d$ state of the Pd or Rh within almost the same energy spread. This Ge $4sp^3$ state further forms hybrid orbitals with Pd or Rh $4d$ state resulting in a covalent bond between Ge and Pd or Rh [42, 43]. Another noticeable feature observed is the difference between the $4d$ states of the parent atom Pd and the daughter atom Rh in Ge. The Pd 4d sub-states are more localized than the corresponding sub-states of the Rh. This is obvious in terms of the sub-state energy spreads of the metals, which are about 2.0 eV (between 1.3 eV and 3.3 eV below VBM) for Pd and 3.3 eV (between VBM and 3.3 eV below VBM) for Rh. Therefore, as a result of the higher spread of the Rh 4d sub-states, they participate more in the hybridization than the Pd but the Pd 4d sub-states are buried deeper in the valence band and stronger bond than that of the Rh [44, 45]. This was also observed in Si. This observation shows the slight difference between the chemical nature of the parent and the daughter atoms in the host lattice.

It is a well-known deficiency that the DFT-GGA approach largely underestimates the band gap of materials, especially semiconductors. The calculated band gaps for bulk Ge and Si in this work are underestimated as 0.09 eV and 0.64 eV, respectively, which are smaller than the experimental values 0.67 eV and 1.17 eV but close to the calculated values 0.08 eV and 0.65 eV of other authors [36, 46]. However, this underestimation of band gaps does not invalidate the fine structure of the bands and the computation of the observables, which depend only on the ground state properties for instance, the EFG and the atomic coordinates [9].

Table 2: Lattice relaxations of the impurity atom (Pd or Rh) and the six NN atoms in the complex calculated as a percentage of the distance between two nearest-neighbour (NN) atoms of an unrelaxed host (Ge or Si). The relaxations are in units of NN%.

| Charge | Si-Pd | | Si-Rh | | Ge-Pd | | Ge-Rh | |
|---|---|---|---|---|---|---|---|---|
| | Pd | Si | Rh | Si | Pd | Ge | Rh | Ge |
| 0 | 29 | 6 | 22 | 5 | 31 | 13 | 27 | 12 |
| -1 | 49 | 10 | 30 | 9 | 50 | 13 | 49 | 17 |
| -2 | 50 | 11 | 48 | 14 | 49 | 13 | 50 | 17 |

Table 3: Binding energies $E_b$ (eV) for impurity-vacancy pairs in Ge and Si.

| defect complex | Ge $E_b$ (eV) | Si $E_b$ (eV) |
|---|---|---|
| $Pd_{SS}$-V | -0.37 | -1.38 |
| V-$Pd_{BI}$-V | -0.85 | -1.84 |
| $Rh_{SS}$-V | +0.23 | -0.58 |
| V-$Rh_{BI}$-V | -0.44 | -1.15 |

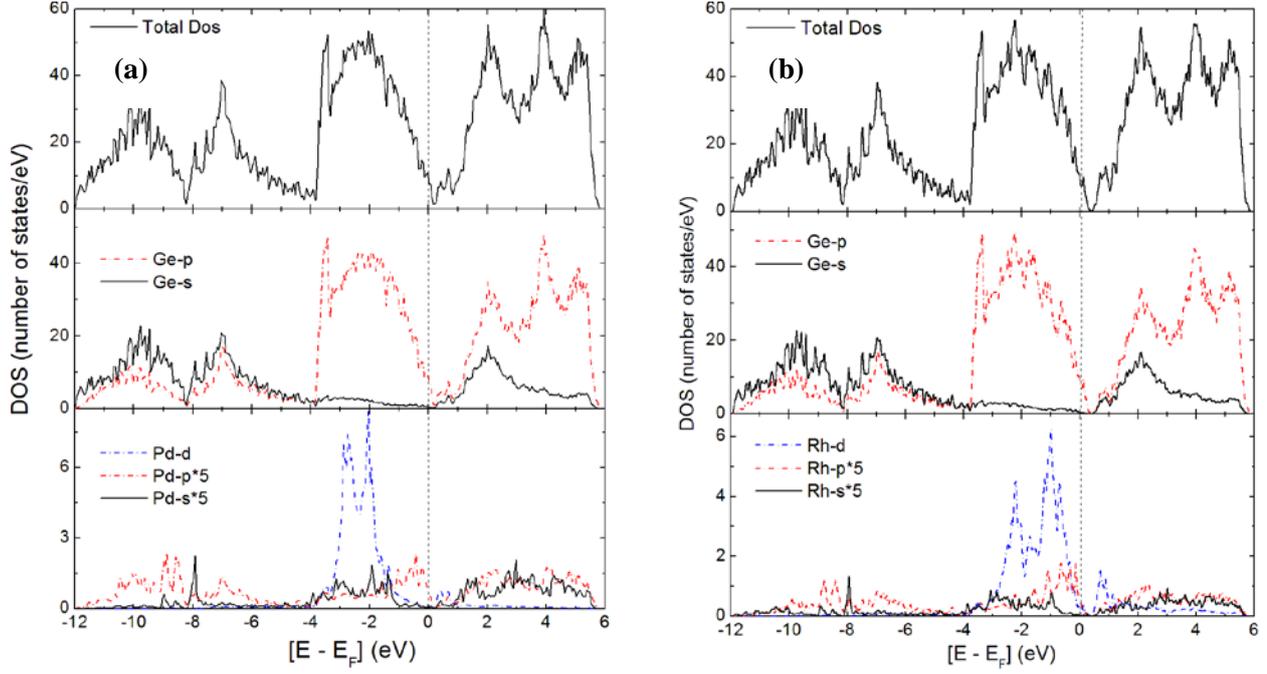

**Fig. 6.** (Color online) Total and partial density of states (PDOS) for impurity-dope Ge: (a) the Pd-doped Ge; (b) the Rh-doped. The Fermi level is located at the zero of the energy scale.

5.2. Binding energies and charge-density analysis

The interactions of point defects with each other in a crystal lattice can be evaluated by calculating the binding energy $E_b$ acting between them. The binding energy is defined as

$$E_b = E_{defect\ cluster} - \sum E_{isolated\ defects}. \qquad (9)$$

According to this equation, negative binding energies imply that a complex is energetically favourable with respect to its constituent isolated components [47].

Therefore, the binding energy of a PdV pair in Ge is given by

$$E_b(\text{PdVGe}_{n-2}) = E(\text{PdVGe}_{n-2}) - E(\text{PdGe}_{n-1}) - E(\text{VGe}_{n-1}) + E(\text{Ge}_n) \qquad (10)$$

where $n$ is the number of atoms in the supercell, $E(\text{PdVGe}_{n-2})$ is the energy of a supercell containing ($n$–2) Ge atoms and V is a vacancy. $E(\text{Ge}_n)$ is the energy of a supercell containing $n$ atoms of Ge, $E(\text{VGe}_{n-1})$ is the energy of a supercell containing one defect and $E(\text{PdGe}_{n-1})$ is the energy of a supercell containing one metal impurity. This definition of binding energy is also used to evaluate the binding energy of the same complex in Si for comparison with Ge.

TM impurities introduce deep-level states in elemental semiconductors due to the formation of impurity-defect complexes [4]. Therefore, two main contributions to the binding energy between Pd and vacancy in these semiconductors are: (*a*) the relaxation of impurity metals and the vacancy; and (*b*) the contribution arising from electronic effects. In this work, only the neutrally charged defect complexes were investigated for the computation of the binding energies.

The predicted binding energies of Pd-defect pairs in Ge and Si calculated here according to Eq. (10) are summarized in table 3. Our calculations show that the two Pd-defect configurations are energetically favourable. However, we observed that the V-Pd$_{BI}$-V configuration was more stable than the Pd$_{SS}$-V configuration by about 0.47 eV in both Ge and Si. This is in agreement with the EFG calculations. To further study the influence of the transmutation of the $^{100}$Pd nucleus to its daughter $^{100}$Rh nucleus on the complex, similar calculations were performed for Rh-defect complexes. The binding energies of the Rh-defect complexes were also calculated, and are shown in table 3. As before, the V-Rh$_{BI}$-V configuration is more stable than the Rh$_{SS}$-V configuration by about 0.67 eV in Ge and 0.57 eV in Si. On an impurity basis, the V-Rh$_{BI}$-V configuration is less stable than V-Pd$_{BI}$-V configuration by about 0.41 eV in Ge and 0.59 eV in Si. This is in agreement with the PDOS analysis in the preceding sub-section. We deduced from the binding-energy calculations that the split-vacancy (V-impurity-V) configuration is more energetically favourable than the substitutional impurity-V configuration in both Ge and Si for both TM impurities, Pd and Rh.

## 6. Summary and conclusions

Our TDPAC measurements have established that Pd-V complexes are present in undoped Ge. This complex was initially interpreted as a Pd atom on a substitutional site pairing with a nearest-neighbour vacancy in n-type doped Si. However, our DFT calculations of Pd atoms in Ge and Si, and the calculations coupled with the experimental measurements (e.g. emission channelling) of other researchers for TM impurities, indicate that Pd atoms in fact occupy bond-centred interstitial (BI) positions in Ge and Si. Thus, the DFT calculations have been used to predict a split-vacancy configuration, with the Pd on a BI site having a nearest-neighbour semi-vacancy on both sides (V-$Pd_{BI}$-V) with negative charge states. In addition, the binding-energy calculations show that the split-vacancy (V-impurity-V) configuration is more energetically favourable than the impurity-V configuration in both Ge and Si. Finally, the calculations indicate that the transmutation of Pd to Rh during the TDPAC measurements has negligible effects on the lattice-site location of Rh.

**Acknowledgments**

We thank Dr Davide Ceresoli (one of the developers of QE-GIPAW code) for his invaluable advice and technical support on the DFT calculations of the EFG's. The DFT calculations were supported by the NCI National Facility at the Australian National University. The Centre for Energy Research and Development, Obafemi Awolowo University, Ile-Ife, is acknowledged for granting one of the authors (AAA) study leave for this work.